\documentclass[manuscript]{aastex631}

\usepackage{natbib}

\shorttitle{Spectrum: Empowering Equitable Excellence}
\shortauthors{Latouf et al.}

\begin{document}
\title{Effective \& Ethical Mentorship in Physics and Astronomy through Grassroots Organizations}

\author[0000-0001-8079-1882]{Natasha Latouf}
\altaffiliation{NSF Graduate Research Fellow, 2415 Eisenhower Ave, Alexandria, VA 22314}
\affiliation{Department of Physics and Astronomy, George Mason University, 4400 University Drive MS 3F3, Fairfax, VA, 22030, USA}
\affiliation{NASA Goddard Space Flight Center, 8800 Greenbelt Road, Greenbelt, MD 20771, USA}
\affiliation{Sellers Exoplanents Environment Collaboration, 8800 Greenbelt Road, Greenbelt, MD 20771, USA}

\author[0000-0002-6454-861X]{Emma Schwartzman}
\affiliation{Department of Physics and Astronomy, George Mason University, 4400 University Drive MS 3F3, Fairfax, VA, 22030, USA}
\affiliation{U.S. Naval Research Laboratory, 4555 Overlook Ave SW, Washington, DC 20375, USA}

\author[0000-0002-0913-3729]{Jeffrey McKaig}
\affiliation{Department of Physics and Astronomy, George Mason University, 4400 University Drive MS 3F3, Fairfax, VA, 22030, USA}

\author[0000-0003-3152-4328]{Sara Doan}
\affiliation{Department of Physics and Astronomy, George Mason University, 4400 University Drive MS 3F3, Fairfax, VA, 22030, USA}

\author{Joseph Weingartner}
\affiliation{Department of Physics and Astronomy, George Mason University, 4400 University Drive MS 3F3, Fairfax, VA, 22030, USA}

\correspondingauthor{Natasha Latouf}
\email{nlatouf@gmu.edu, natasha.m.latouf@nasa.gov}

\begin{abstract}

Effective and ethical mentorship practices are crucial to improving recruitment and retention especially for historically minoritized groups (HMGs). Spectrum is a diversity, inclusion, equity, and accessibility (DEIA) grassroots organization committed to empowering equitable excellence through sustainable change. By improving transparency and DEIA within the fields of physics and astronomy, we can empower the next generation of diverse scientists and increase field retention. Starting within our home department at George Mason University and moving outwards, we ensure our students leave as advocates for DEIA and AJEDI (access, justice, equity, diversity, and inclusion) through education and mentorship. Spectrum is providing professionally trained peer mentors to aid students in all facets of their academic and personal lives. Although the peer mentoring program existed since the creation of Spectrum in Spring 2020, we have recently developed and implemented a formal mentorship training for both student and faculty mentors thus increasing the quality, trustworthiness, and confidence of our mentors. Using the latest mentorship research available, this training is developed by Spectrum for George Mason University, with the ability to implement the training at any institution.


\end{abstract}

\section{Reducing Barriers in Physics and Astronomy through Mentorship}
\label{sec:import}

It is well established that historical and current systemic barriers have resulted in disproportionately low rates of recruitment and retention for historically minoritized groups (HMGs) in physics and astronomy. Women earn only 21\% and 33\% of the bachelor’s degrees in physics and astronomy respectively, percentages that have remained stagnant for nearly the last decade (2007-2017) despite a 46\% increase in the number of physics degrees awarded in the same time period \citep{ivie19}. While there has been some improvement in these numbers for women in the past few years, they still lag behind the expected values given our population. Students from historically minoritized racial groups, including Black, Latine, and Indigenous students, have seen even less improvement, earning only 16\% of bachelor’s degrees, and 6\% of PhDs in astronomy in 2014, with very little improvement since \citep{jones18}, although astronomy is generally considered to have higher parity rates than physics. It is prudent to note that traditionally, the definition of HMGs includes only Black, Latine, and Indigenous students, thus excluding a large number of other HMGs such as Arabs and Arab Americans. If the definition is expanded to include intersectional identities (i.e. identities that span several different axes across race, ethnicity, gender, etc.), the statistics plummet, in some cases so low so as to be unable to be quantified. For instance, although there were 59,894 PhDs in physics awarded from 1972 to 2017, only 90 of those were awarded to Black women (0.15\% of the total amount) \citep{miller19}. 

For each of these groups and identities, there are four traditionally challenging barriers faced by HMGs: imposter phenomenon, stereotype threat, othering, and microaggressions. One could argue there are additional challenges, including false allies and the overburdening of early career researchers (ECRs). These barriers keep the rates of recruitment and retention low by continuing the cycle of oppression and exclusion in physics and astronomy. Until these barriers are understood and addressed through mentorship and organizational shifts, the rates of HMGs in physics and astronomy will not improve to the levels expected given current population statistics. What follows is an introduction to each barrier, along with the strategies of mentorship that can aid in mitigating the effects of barriers. These descriptions will mirror \citet{markle_supporting_nodate}; for further information, please see \citet{markle_supporting_nodate} and references therein. 

\subsection{Imposter Phenomenon}
Imposter phenomenon is defined as the feeling that one is not truly as capable or intelligent as others perceive them to be, resulting in the self-perception that one is a fraud \citep{clance1978imposter}. Along with a deep-rooted conviction of fraudulence, there is a constant fear of being discovered or exposed by others that they believe ``belong'' in the field. Thus, it is difficult for individuals experiencing imposter phenomenon to be assured that their success is merited by their ability, hard work, and intelligence \citep{harvey1985if, chakraverty2020impostor, feenstra2020contextualizing, markle_supporting_nodate}. While imposter phenomenon has historically been known as imposter \textit{syndrome}, the use of the term syndrome indicates that the fault lies within the individual experiencing the phenomenon (i.e. it is an internal syndrome), and thus there has been a shift to using the term imposter \textit{phenomenon} instead, indicating a fault within the system that aids in the oppression of individuals in the system. Imposter phenomenon is more prevalent among students from HMGs, due to the lack of diverse mentors and role models. The result is an increase in several mental health concerns, such as depression, anxiety, low self-esteem, low self-confidence, perfectionism, self-doubt, procrastination, and self-handicapping \citep{thompson1998attributional, thompson2000impostor, ferrari2006impostor,fraenza2016role, blondeau2018relation, markle_supporting_nodate}. The combination of the worsening mental health and imposter phenomenon looming tends to result in ``spirals,'' wherein a student consistently tries to prove to themselves that they do not belong (e.g., by being too depressed to study well for an exam, and thus performing poorly). The result is what seems to them to be proof that they do not belong, which then furthers the imposter phenomenon and associated mental health concerns, and the cycle continues.

Many studies indicate that having professional or personal relationships with others in academia, as a peer, mentor, colleague, etc., can significantly reduce the prevalence of imposter phenomenon, including among students from HMGs \citep{l_baker_mentor-protege_2014, feenstra2020contextualizing, markle_supporting_nodate}. Therefore, to attempt to reduce impostor phenomenon, individual characteristics and the environment cannot be the only criterion taken into account by a mentoring program; rather, the fit of the mentor and mentee is crucial and there is significant evidence to suggest that the proper fit can reduce imposter phenomenon \citep{sanford2015finding, cohen2019fear, barr2020coping, chakraverty2020impostor}. This mentor-mentee fit needs to account for three facets of identity perception: professional identity, relational identity, and personal identity. These facets of identity indicate the self-perception related to the roles of their academic career, roles within family and friend relationships outside of a professional environment, and the general sense of self and perception of characteristics, respectively \citep{l_baker_mentor-protege_2014}. By arranging mentor-mentee pairings based on their self-perceptions in these three major areas, imposter phenomenon can be reduced by building a sense of kinship and thus forging a deeper level of trust in the relationship. Mentors can assuage worries about belonging with their own experiences, and thus guide their mentees through the feelings of imposter phenomenon. While this does not obliviate imposter phenomenon, it aids students in realizing that others also struggle with belonging and they are not alone. 

\subsection{Stereotype Threat}

The performances of HMGs can be negatively impacted by identity stereotypes when they are engaging in activities that are related to those stereotypes; this phenomena is known as stereotype threat \citep{steele1995stereotype, thomas2018stemming}. \citet{woodcock2012consequences} used a large, diverse sample of HMGs in STEM, and found that stereotype threat was associated with ``scientific disidentification'', i.e., HMGs find themselves belonging to groups they do not wish to belong to. This leads to a decline in the affected student's persistence toward entering a STEM career. There is extensive literature that demonstrates that the development of a physics identity is crucial to the student persisting in STEM. \citet{hazari10} developed a model to better understand two key aspects of a physics student's identity:
\begin{enumerate}
    \item How the ability to conceptualize physics careers beyond academia impacts female students' physics identity and sense of belonging (especially when considering students who take non-traditional paths in academia or career trajectories).
    \item How physics leadership opportunities contributes to building physics identity and sense of belonging in the field. 
\end{enumerate}

Mentors are crucial for providing the role model and support that many students need, whether in the form of help conceptualizing physics careers beyond academic or encouragement in the pursuit of leadership opportunities. However, stereotype threat is also persistent for HMGs at the faculty level, with similar effects to those noted at the student level \citep{casad2016addressing}. Since these effects are so prevalent, less faculty positions are filled by HMGs, and those in place are reluctant to pursue leadership roles that would make them accessible to HMG students who would benefit from their advocacy. 

While many studies have sought prescriptions to reduce stereotype threat, several findings are not replicable (see \citet{markle_supporting_nodate} and references therein). This is potentially due to the fact that these studies and affirmation exercises are treating the symptoms of a systemic issue, while overarching transformations in the social and academic systems in power are what is needed to treat the disease itself. Culturally responsive mentoring is defined as mentorship wherein the mentor acknowledges their privilege and uses the cultural knowledge and frames of reference of HMGs to make learning more relevant to and effective for them \citep{byars-winston_science_2019}. Culturally responsive mentoring has been shown to effectively combat the effects of stereotype threat on HMGs. This is a long-term, consistent, and institutionalized method to ensure that HMGs at all career stages that partake receive the affirmation of identity and value they need, thus further building and bolstering physics identity  \citep{mondisa_social_2015, san_miguel_successful_2015, byars-winston_science_2019}.  This requires the mentor to be comfortable with being uncomfortable, and assumes constant growth \citep{harris_advocate-mentoring_2019}. There are recent studies that support culturally responsive mentorship as an effective mentorship method, based on student interviews and self-perception tests, thus providing preliminary feedback that students benefit from interactions with mentors who they can see themselves in, i.e. ``looks like me'' \citep{thomas2018stemming}. Looking specifically at undergraduate HMGs, those who received culturally responsive mentoring felt greater confidence as researchers and were more committed to continuing through a graduate degree \citep{haeger2016mentoring}. This builds on prior research wherein students named their role models and mentors as their primary inspiration to pursue STEM degrees and persist in the field through any challenges, thus further cementing the importance of culturally responsive mentoring \citep{meador2018examining}. Having a network of peers can also be an effective strategy for mitigating stereotype threat as found in \citet{block_contending_2011}, as well as other barriers. 

\subsection{Othering}

Low levels of social connection and ``belongingness'' (i.e., the perception of acceptance, connection, social support, feelings of mattering, respect, and being valued) can result in a reduction of interest in STEM and low retention in students from HMGs \citep{stachl2020sense}. On the other hand, having a strong connection to one's peers built on positive interactions can result in higher rates of academic achievement, persistence in field, and retention in STEM \citep{zaniewski2016increasing, ito2018factors, kim2018science, robnett2019form}. There are several factors that lead students from HMGs to feel isolated, especially at Predominantly White Institutions (PWIs), thus leading to a social separation from their peers. These include a poor campus climate, racist/sexist/xenophobic/other hateful interactions with White peers or faculty members, and having a very strong connection to their racial identity that is marginalized or bullied at their institution \citep{brown1990racism, locks_extending_2008, thelamour2019we}. These feelings of isolation are a powerful impediment to fostering the culturally inclusive environment at the institutional level (that begins at the departmental level) necessary for both the success of STEM students in HMGs \citep{fisher2019structure} and increasing retention in students from HMGs in STEM \citep{johnson2020culturally}.

Mentoring relationships can drastically improve a mentee's sense of belonging \citep{chan_mentoring_2008, stolle-mcallister_meyerhoff_2011, apriceno_mentorship_2020}. This is not specific to just students from HMGs; \citet{apriceno_mentorship_2020} found that students from all groups who have an engaged mentor early in their career develop greater senses of belonging and self-confidence in academia and in STEM. Engaged mentorship does not have to only come from a senior career member in STEM; mentoring that is received from peers has a profound impact on a student's sense of belonging. Connections between students who are closer in career stage (ideally with similar backgrounds or interests) can both emphasize and validate their struggles within STEM and thus aid in the development of their STEM/physics identities \citep{maton2000african, inzlicht2006environments, zaniewski2016increasing}. Moreover, studies specifically within physics have shown that students who participate as mentors in constellation mentorship programs (programs that connect mentors and mentees from multiple career stages) bolster their own physics identity, as well as aid in striking a sustainable balance between their science and their commitment to the field, leading to increased long-term satisfaction \citep{godwin16, pando22}. By combining the forces of student mentorship with positive senior career researcher (SCR) mentors (i.e. emphasizing constellation mentorship models), the chances of success of students from HMGs increases significantly. Student mentors will naturally have more available time and a more similar perspective, while the SCRs will be able to provide a better ``big picture'' view of the field and help guide a mentee through career transition periods with greater efficacy.  

\subsection{Microaggressions}

The inability for a student to relate to a mentor that looks like them or can empathize with them on multiple axes can intensify imposter phenomena and, in many cases, can lead to the student being pushed out of the field. These feelings are exacerbated by the constant microaggressions faced by HMGs on the basis of race, gender, disability, sexual orientation, and other identity markers \citep{tatum2000whoami,swim2001everyday,evans2002experiences, alleyne2004black, sue2010microaggressions, young_hierarchical_2015}.

The term ``microaggression'' describes a covert, unconscious, subtly destructive phenomenon related to individualized experiences with othering in society \citep{young_hierarchical_2015}. Although used as an umbrella term, multiple different forms of aggression can occur in the workplace environment that can be defined as ``microaggressions''. \citet{sue2010microaggressions} researched microaggressions in everyday life and developed a typology of microaggressions to better understand their occurrence and impact. There are three distinct types of microaggressions: microinsults, microinvalidation, and microassault. Microinsults convey rudeness or insensitivity through behaviors, actions, or verbal remarks in an attempt to diminish, exclude, or negate the experienced reality of HMGs. Microinvalidations feature the use of actions to exclude, negate, or nullify the experienced reality of HMGs (i.e. more insidiously excluding a person to other them). Finally, microassaults are explicit put-downs to hurt the victim, e.g. outright racism \citep{sue2010microaggressions, young_hierarchical_2015}. Microaggressions are used to create and maintain a hostile environment for HMGs, leading to a higher probability of burn out or field departure. Microaggressions cut deepest when perpetuated by those who might be considered allies. \citet{harris_advocate-mentoring_2019} noted that mentoring can combat the effects of microaggressions, but all rely strongly on the connection between the mentor and mentee. To combat microaggressions, the pair must:

\begin{enumerate}
    \item Openly discuss cultural responsibility and acknowledge different experiences due to race or ethnicity.
    \item Openly discuss what the mentee needs specifically in relation to race- or ethnicity- based challenges.
    \item Openly discuss the mentor's plans for addressing social injustices and supporting the mentee.
    \item Specifically the mentor must utilize positive micromessaging in order to continuously affirm competency and foster growth (e.g. ''great question", ''that's a clever solution, but why don't we try it this way", etc.)
\end{enumerate}

This open communication is crucial, as it ensures that the mentor/mentee pair are in agreement on fundamental growth points and understand the goals that they are each working towards. Especially when concerning topics relating to minoritization, proper training on effective communication can make the difference in a successful or harmful relationship.

\subsection{False Allies and Overburdening ECRs}
Although many people may call themselves an ally to the cause, and perpetuate an image of a helpful advocate, in many cases, well-intentioned white male physicists also aid in perpetuating inequity in the field both consciously and unconsciously. Many have internalized sexism or racism, and justify their inaction using three main themes found by \citet{dancy22}. For further information on each of the three themes listed below, as well as an in depth discussion of the phenomena, please see \citet{dancy22}:
\begin{enumerate}
    \item Physical Distancing: By persisting in the belief that racism and sexism do not happen in their classrooms, departments, research groups, institutions, or the field of physics, inaction is justified.
    \item The Problem is Too Large: By claiming that the problem is on too large a scale for one person to have an impact, i.e. blaming the K-12 system, historical legacy of sexism and racism, socioeconomic dynamics, or differential expectations of parenting, inaction is justified.
    \item I am Helpless to Act: By refusing to see the issues, and thus using that as justification for inaction, or claiming that action would create negative consequences for the harrasser, they claim that they are not capable of acting and thus inaction is justified.
\end{enumerate}

By presenting themselves as powerless, yet still using the ``ally" moniker, many physicists in positions of power can perpetuate inequity and microaggressions within their field and institution while maintaining a white-savior complex. The burden of creating a more equitable field falls then on the shoulders of historically minoritized and early career groups such as undergraduate students, graduate students, and post-doctoral researchers. Many organizations have formed to push the field forward, relying on early career researchers (ECRs) volunteering to make a change. Groups such as Black in Astro\footnote{https://www.blackinastro.com/} \citep{walker22}, and League of Underrepresented Minoritized Astronomers (LUMA)\footnote{https://www.lumamentoring.com/} are working to combat the state of the field by providing a community for HMGs, although both are created and led by ECRs. These organizations have resulted in an increase in acts of service, done voluntarily and designed to improve the state of the field. However, they have faced opposition at every turn, including while trying to implement mentorship practices. One such organization was developed internally at George Mason University. The development of these resource hubs and community groups for students is crucial to improving recruitment and retention. However, the most effective method of retention lies in ethical mentorship practices that address the common barriers to STEM while specifically aiding HMGs \citep{byars-winston_science_2019, markle_supporting_nodate}.

\section{Applying Recommended Best Practices for Mentorship}
\label{sec:practices}

\begin{figure*}[]
\centering
\includegraphics[scale=0.3]{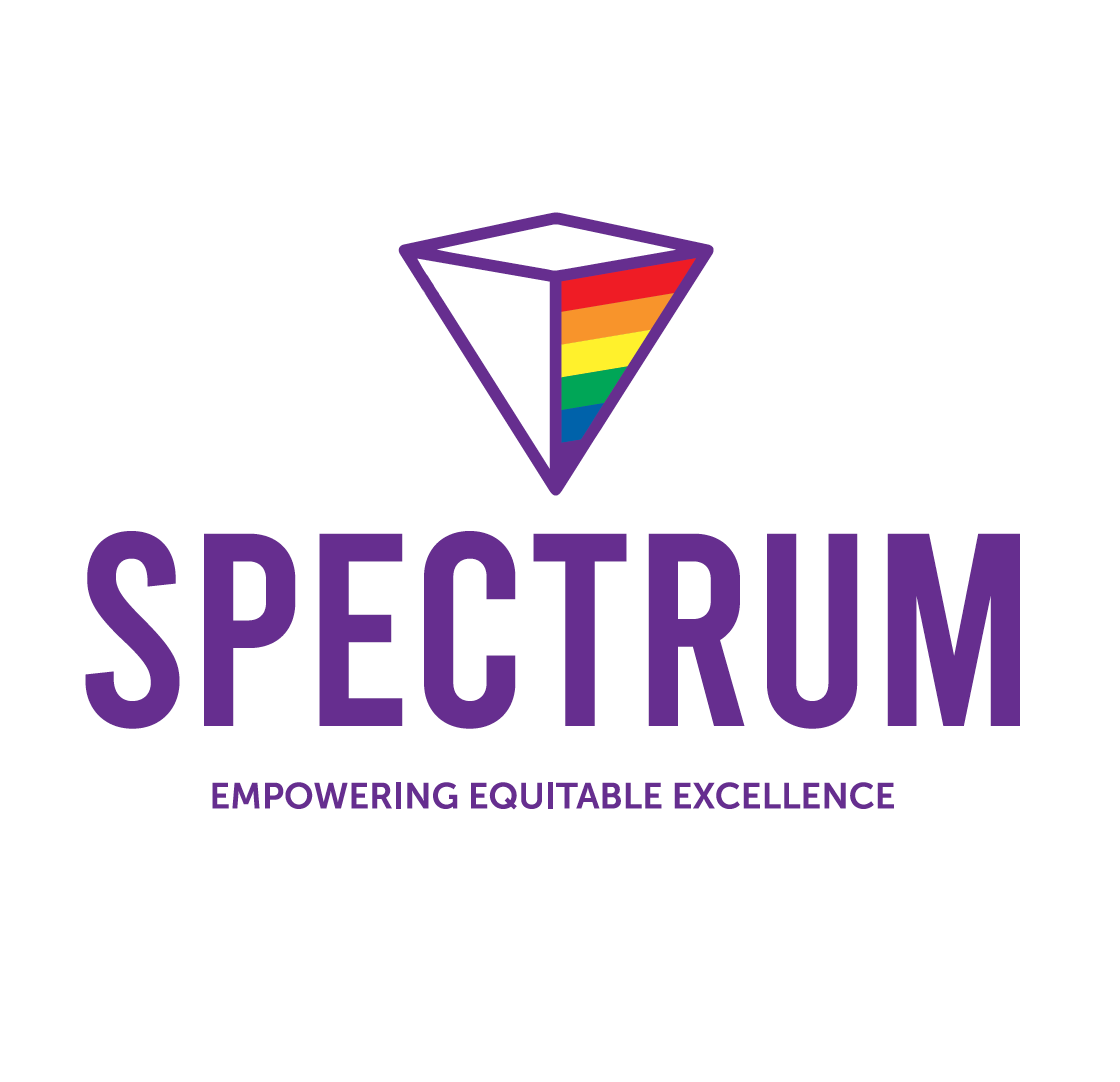}
\caption{Spectrum's designed logo with tagline.}
\label{fig:logo}
\end{figure*} 

\subsection{Spectrum: Empowering Equitable Excellence}
\label{sec:spectrum}

In February of 2020, four physics and astronomy students from HMGs, including author Natasha Latouf, were approached by a faculty member and asked to develop the first Code of Professional Conduct for the George Mason University (GMU) Department of Physics and Astronomy. Previous to this request, GMU Department of Physics and Astronomy had had no such code, exposing students and faculty from HMGs to suffer from the effects of sexism, racism, retaliation, etc. The Code of Professional Conduct was drafted and adopted into the GMU Department of Physics and Astronomy in May of 2020. Spectrum\footnote{https://gmuspectrum.squarespace.com/} was officially founded shortly thereafter. 

Spectrum is a student-created and student-run access, justice, equity, diversity, and inclusion (AJEDI) grassroots advocacy organization in the GMU Department of Physics and Astronomy. The primary goal of this organization is to drive sustainable change to improve transparency, diversity, equity, and inclusion within the fields of physics and astronomy, starting within our home department. To this end, a significant responsibility that Spectrum has taken on is to ensure that our students not only leave our department as well-rounded scientists, but also as strong ambassadors and advocates for AJEDI, wherever their path leads them next. We believe in empowering equitable excellence within physics and astronomy through education and mentorship. 

The process of building Spectrum was arduous, having faced opposition within the department and difficulties due to the COVID-19 pandemic, including forming a community in a virtual environment. Spectrum was also unfunded for the first two years of its existence, relying solely on the volunteering of students. Despite these many barriers, the founders, leaders, and members of Spectrum persevered, speaking out and conducting numerous events.  

Over the last four years, Spectrum has undertaken several initiatives to improve the status of the field. Spectrum co-founders worked with department faculty and HR staff to develop and present a Bias and Inclusion training, hosted as a part of the GMU departmental colloquia series. This training garnered over 60 attendees, becoming the most well-attended seminar of the semester to date. We also founded the Professional Development Seminar Series, which focused on providing assistance in both traditional (i.e. resume writing, science publication, research opportunities, etc.) and non-traditional (mentorship, recognizing red flags in academia, imposter phenomena, etc.) professional development. This included hosting invited guest speakers with specialized expertise in topics of interest while also developing and presenting talks ourselves to give a near-peer perspective on strategies to be successful in a STEM career. This seminar series was weekly, moving to bi-weekly, and all talks were recorded and uploaded to our YouTube channel\footnote{https://www.youtube.com/channel/UCjkWQLgKgcW9YhtoFOibyOg} to allow watchers to avoid Zoom fatigue and learn on their personal schedule \citep{elbogen2022national}. We also hosted several extended events and workshops, including Pathways to Government, a Virtual Teaching and Learning Open Discussion, Personal Statements and Presenting Workshop, Recognizing and Overcoming Imposter Syndrome, etc. 

To continue the learning experience for all members of the department, Spectrum also undertook the task of building a website filled with resources and educational opportunities. This includes a monthly themed webpage for DEIA in honor of each diversity celebration month spotlighting at least three scientists within each relevant month. The website provides an avenue of education for students to learn from the past, and for HMG students to see themselves represented. The website also features an extensive resources tab, with focus on Skill Building, Career and Grad School Planning, Mental Health and Time Management, DEIA in STEM and Society, Science Outreach, etc. to provide students with a central location to access any manner of resources, and break down traditionally gatekept knowledge barriers.

In an effort to build community with other STEM fields and more public audiences, Spectrum has undertaken a number of social activities. Spectrum co-hosted a Minorities in STEM mixer, collaborating with six other minority in STEM groups across GMU's College of Science in order to allow members of each organization to mingle and make valuable connections across STEM fields. Leaders also revitalized the Graduate Research Assistant Office, investing personal funds and time to paint walls, purchase and build furniture, and provide decor, in order to encourage a community among the graduate students in the department. To increase education and interest external to the GMU Department, Spectrum participated in the local Fairfax Fall Festival to shine a spotlight on our mentoring and outreach initiatives to $ge$5,000 attendees. Using this platform, we were able to inform prospective students about our open source resources, increase engagement in science by providing an ``Ask a Scientist'' section, and encouraged participation from younger children interested in growing their scientific passion.

\subsection{Spectrum Mentorship Program}
\label{sec:specmentor}

Despite all of the other, very useful efforts of Spectrum, the most important has been the development of a premier mentorship program within the physics and astronomy department. Our mentoring program utilizes the theoretical framework of a physics student’s identity, described above and by \citet{hazari10}, and was developed after the constellation mentorship model. After an extensive intake process asking both personality and mentoring needs-based questions, we match each mentee to two mentors based on factors that will provide the most optimal personality match, such as most active time of day and personal and professional interests. These matches are designed to best reinforce and mold their student identities. By having two mentors per mentee, we increase the diversity of perspectives each mentee is exposed to, and reduce the time commitment of each individual mentor. Additionally, \citet{dennehy17} demonstrated that when female peer mentors are introduced to undergraduate female students in career transition periods (e.g. deciding on graduate school, or industry, etc.), there was a marked increase in belonging and self-efficacy, resulting in long-term higher retention. Similar results were found for race and ethnicity \citep{trawick21}. As such, we strive to match each mentee with at least one mentor of the same gender and ethnic background, if available, and provide avenues to contact other mentors of their same gender and ethnic background to further increase the likelihood of retention. This aids in overcoming the barriers presented by othering, microaggressions, and a lack of a sense of belonging. We have also appointed faculty members to the Spectrum Advisory Board, who are available to provide guidance upon request to Spectrum leadership, and as appointed, trusted faculty mentors for students who are looking for the faculty perspective. Our mentoring program currently services approximately approximately 30\% of the department (as of Spring 2024).  

The primary source of training for mentors until Fall 2023 has been through relatively informal education, through a compilation of resources discussing best practices distributed through Spectrum’s website, with targeted discussions on topics such as imposter phenomena, self-advocacy, stress management, and effective communication. Our faculty mentoring committee’s training has been equally informal. While some of our faculty mentors have had the opportunity to take formal training in mentorship, others work primarily off of their lived experience, best practices from the literature, and Spectrum events, such as the Bias and Inclusion training departmental seminar.

As our mentorship program has grown, it became necessary to formalize and standardize our training procedures to ensure that our mentors are well equipped with the proper tools to best support their mentees, while also knowing how to take care of themselves while advising students in potentially sensitive situations. As described in \citet{markle_supporting_nodate}, mentor training is a crucial component to an effective, holistic, non-dyadic mentorship model. A holistic, non-dyadic mentorship model focuses on network-based mentorship as an alternative to traditional mentorship (i.e., a dyadic model) wherein there is only the connection between mentor and mentee. Holistic, non-dyadic mentorship enables mentees to also be mentors and have a peer group at each level, building a continuing web of aid and mentorship (e.g., a faculty member (or members) mentors a grad student, who mentors an undergrad, while the grad student and undergrad are connected to their peer communities). Studies have found that students who had significant mentoring relationships with multiple levels of expertise (faculty member, post doctoral fellow, graduate student) reported better productivity, stronger scientific identity, and a higher likelihood of pursuing higher education \citep{aikens_race_2017, joshi_direct_2019}. Similarly, \citet{collings_impact_2014} found that undergrads who participated in peer mentoring (i.e. being mentored by others in their peer group) were less likely to leave their university. This cascading effect of mentorship has been proven time and time again \citep{byars-winston_science_2019, markle_supporting_nodate}, illustrating that a holistic model is likely to be the most effective for recruitment and retention. 

\begin{figure*}[]
\centering
\includegraphics{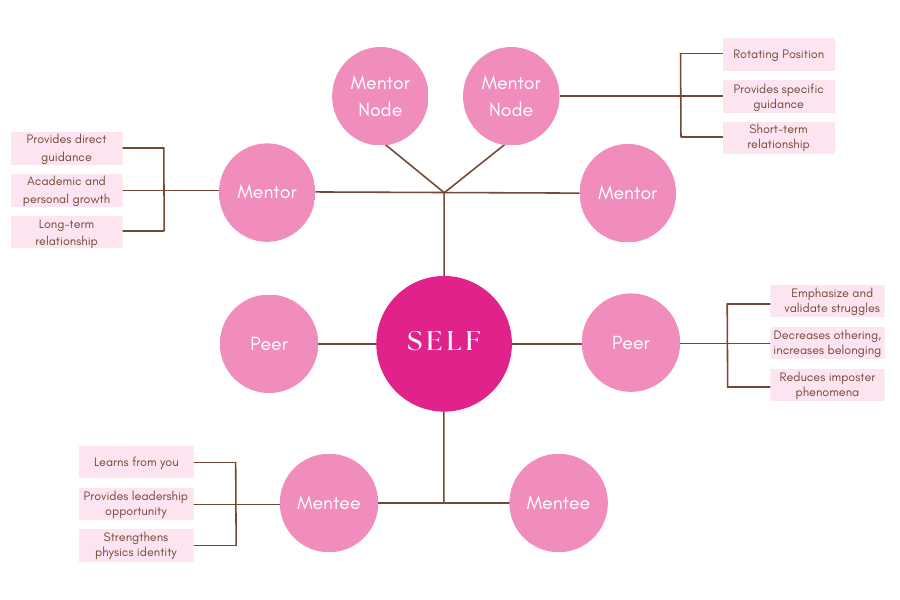}
\caption{We present a model of a holistic, non-dyadic mentorship model (constellation mentorship) that includes all of the available mentorship positions: mentor node, mentor, peer, and mentee, surrounding the central self. Each position is described in detail, and one can imagine this same model repeating per individual.}
\label{fig:mentorshipmodel}
\end{figure*}

\citet{markle_supporting_nodate} and \citet{byars-winston_science_2019} both present figures (Figures 1 and 4-1d, respectively) describing the set-up of a holistic non-dyadic mentorship model. In \citet{markle_supporting_nodate}, the graduate students all look to one faculty member as a mentor, although it is implied that there should be more than one faculty member in the model; there is also no placement for ''mentor nodes'', which are rotating, short-term positions that are filled by mentors for specific projects or academic transition periods. In \citet{byars-winston_science_2019}, they describe the differences between each mentorship model with panel d presenting a network model; however, it is a combination of panels c and d that best describe a constellation mentorship model, as panel d does not feature a location for the multi-level mentorship or equivalent-level peers. Extending this work, Figure~\ref{fig:mentorshipmodel} describes a robust holistic, non-dyadic mentorship model, combining elements from both of the excellent figures presented above. It includes the concept of ``mentor nodes'', mentor slots which are long-term relationships that emphasize both academic and personal growth, peer slots to aid in reducing othering and increasing belonging, and mentee slots to assist with developing leadership and physics identity. Each position is described, and connected to the self in the central position. It is after this model that we develop our training materials and mentorship model within Spectrum. This mentorship model requires an effective strategic plan to thrive, and that indicates the need for training.

However, there is typically little to no training provided to mentors that centers around research-backed best-practices in mentorship\citep{pfund_merits_2006}. It is generally accepted that a one-time short training is insufficient for building durable skills, and that regularly occurring meetings or sessions over time are more effective for imbuing critical points of mentorship practices. However, inplementing these changes will require a greatly reconfigured mentorship training. Given that the physics and astronomy department at GMU has had no mentorship training in the past, the introduction of smaller steps is more likely to engage busy faculty and students alike.  

\subsubsection{Mentorship Training}
\label{mentortraining}

To ensure that all members are working from similar baselines of knowledge, we were awarded an American Astronomical Society Education and Professional Development Mini-Grant (AAS EPD) in Summer 2022 to develop a mentorship training workshop, and the training was implemented in Fall 2023. The workshops were designed based on lessons learned from the past years of Spectrum mentoring, topics requested by our current and former mentors, and the most recent research in effective mentorship practices \citep[Appendix N,][]{decadal20}. The training was split into two parts, designed for faculty and students to take separately, so as to encourage a safe space for growth without an intrinsic power dynamic. In addition to all the topics covered in the SCR training, the training for the ECRs includes topics to protect students, such as recognizing academic red flags, entering into an effective mentorship relationship, and improving communication skills. The workshops both emphasize the five core attributes of effective mentorship developed by \citet{pfund_defining_2016}: research, interpersonal, psycho-social/career, cultural responsibility, and sponsorship. Each of these areas is presented and explained, along with the barriers presented above and how each attribute can aid in mitigating the barriers. We also encourage the mentors to think about their own behavior, and how they could be contributing to either the barriers or the attributes, as well as how to improve and grow. 

As described by \citet{flaxman1988youth}, mentorship can be separated into two categories: planned and natural. Planned mentorship is conducted through structured programs, wherein mentors and mentees are carefully paired through a formal process (such as Spectrum's program). This is the most traditionally used form of mentorship. In contrast, natural mentorship is conducted through friendship, teaching, etc. and is neither formally assigned nor carefully paired. These differences are crucial to the application of mentorship, and by educating the next generation of mentors on these differences, we will supplement understanding in mentorship expectations and increase the many positive impacts of effective mentorship. While mentorship is sometimes viewed as a position one can appoint themselves to (i.e. declaring oneself a mentor), the title of ``mentor'' should be bestowed either by the mentee (natural) or by programmatic appointment (planned). Additionally, there are distinct and important differences between the roles of faculty members, advisors, and mentors in student fostering. Faculty members are not required to mentor all students, but can serve as mentors in multiple cases. Advisors are also not required to mentor their students, however they are obligated to develop the research and academic careers of their charges. Mentors are responsible for the personal (e.g. integrity, responsibility, etc.) \textit{and} scientific development of their mentees, within any boundaries agreed upon in the mentor/mentee pair, and should drive open communication and guidance. One of the most crucial roles for mentors is to provide culturally-responsible mentorship. Mentors should not be allies, but rather they should be \textbf{advocates} for students from HMGs. Advocacy is a far more powerful role which utilizes the privilege of the SCR to plead for or on behalf of students from HMGs, and create environments wherein these students can thrive \citep{harris_advocate-mentoring_2019}. Research has suggested that race and ethnicity has a marked impact on mentoring relationships, resulting in the relationships forming more slowly, if at all, in the case of a mixed-race mentorship \citep{pruitt_discrimination_1985}. For an advocacy-based, culturally responsible mentorship relationship to stand a higher chance of success, the mentor must be educated or seek education on their students' marginalized identities and experiences, be comfortable with facing discomfort and their own prejudices in order to unlearn them, and recognize the benefits and privileges they have experienced due to systemic oppression \citep{kendall2003}. By doing this, the mentor will learn and practice advocacy and mentorship from a responsible standpoint, while wielding their privilege as a tool to better the field. To be clear, these mentors will still make mistakes; advocacy and cultural responsibility does not require perfection, but rather participants who will actively learn and from their mistakes \citep{kendall2003}. 

Our training broke down the roles and misconceptions for SCRs, and then built up a healthy understanding of the role of mentor, a mastery of the boundaries involved in each role, and a perception of their own comfort level using the research described herein. Not only did this enable better ECR-SCR relations, but the same awareness will help ECRs understand their own power in self-appointing a mentor. The ECR training included all the same lessons, with the addition of recognizing red flags in an mentoring or advising relationship as well as additional information on open dialogues. Both training sessions included a broad overview of peer mentorship: defining what a ``good'' mentor is, setting proper expectations, understanding healthy communication in mentorship, and providing resources that include accessible, baseline definitions on a variety of AJEDI topics. 

Although AJEDI is frequently lauded as an important aspect of field development, one of the largest barriers in AJEDI education is misinformation, compounded by a lack of unified and trusted resources. For many people at all career levels, there is an ingrained misunderstanding of what AJEDI is, how one can be an effective ally or advocate, or even where to begin unlearning fundamental biases. Correcting this misunderstanding has always been crucial for ethical progress in the field, but more recently has become important for accessing funding. For example, all NASA and NSF-funded proposals now require an Inclusion Plan that will soon be graded as part of the overall proposal. This will force a significant portion of researchers in the field (many of whom are AAS members) to upgrade their understanding of AJEDI topics. In the process, they will inevitably find that the existing materials are insufficient, and will resort to superficial language and vague descriptions. Our training materials help support education in AJEDI topics such as anti-racism, oppression systems, etc. that are crucial for culturally responsible advocacy mentorship. 

Additionally, the training delved into a discussion of the bystander intervention, defined as an individual or group actively addressing a situation they deem problematic that they are observing, and what doing so could mean for an ECR or peer. We recommended seeking out a bystander intervention training, which is frequently offered through a variety of academic institutions. Our workshops also included role-play case studies in which debriefs were used to teach mastering reacting, reporting structure, etc. in difficult situations while in a controlled environment. These role-plays were separated into group conflicts (e.g. student group conflicts), faculty conflicts, crossing boundaries, and work-life balance. This was replicated for the mentee and mentor training, however, the scenarios were altered from student point of view to faculty point of view. These case studies were particularly well received and experienced increased engagement. 

Most importantly, the training provided extensive resources to both the mentors and mentees - including the literature upon which the training was based - so as to continue their education, further their training opportunities, and a mentoring compact. A mentoring compact is an agreement established between the mentor and mentee, including individual responsibilities, possible consequences, and joint goals; it is explicitly a \textit{compact} and not a \textit{contract} since there are no legal implications or methods to uphold the agreement, but rather a good faith agreement to ensure that the boundaries of the mentoring relationship are discussed. This mentoring compact is meant to encourage open dialogue between mentors and mentees by providing the boundaries and opened-ended questions that drive relationship progress (e.g. setting goals, expectations, accountability, etc.). By providing the compact, we remove the apprehension of broaching a ``big picture'' conversation, and decrease the chance of missing an important topic. The mentoring compact and case studies have been provided in Section~\ref{sec:appendix}. A Spectrum starter kit that contains the entire mentoring training as well as all necessary items to start a Spectrum sister organization are available.\footnote{https://drive.google.com/drive/folders/1Bg9ZV9vmdGg6ht9xeuRsexRRu647KzZi?usp=sharing}

\section{Summary, Conclusions, \& Future Works}

There exists a wealth of literature on mentorship research regarding an effective, ethical, culturally responsible, and advocacy-based mentorship program. This paper presents a high-level summary that can be used to drive development of such a mentorship program. Spectrum has developed and tended to such a mentoring program, giving students a space to seek out peer mentorship that is guided by research. This space is safe for students, encourages open dialogue, and breaks down barriers that have prevented students from succeeding in STEM: imposter phenomenon, stereotype threat, othering, microaggressions, false allies, and overburdening ECRs. 

The development of Spectrum was difficult, requiring immense effort and perseverance. We believe that although the onus should not be on the students to drive community improvement, they also should not struggle in the effort to participate in improvement. As such, Spectrum has developed a starter kit of information that is a living document, allowing students to start their own sister organizations without the trial and error, and removing the burden of development and research from the next generation of students. This includes a research-backed mentorship training funded by AAS, a description of the mentorship program with the tools needed to develop their own, all logos and design aspects, a description of the professional development lunch talks, as well as description of the leadership structure. The mentorship training materials are also broken down into subsections for faculty/SCRs and student mentors, to allow for a fine-tuned training without the added power dynamics. 

In future works, we will lead the development of mentorship checklists, designed to prompt critical thought on the environment of a research group. We will design two checklists: one aimed at the mentee level, i.e. early career researchers such as students and post-docs, and one aimed at the mentor level, i.e. faculty, civil servants, PIs, etc. These checklists will work in tandem, with similar questions on either side in order to compare results for a future IRB study on the efficacy of the checklist. 

The mentee checklist will aim to invoke thought about the feelings of the student individually (i.e., do they feel safe, valued, heard by their mentor?) and observation of their research environment (i.e. how does the group interact? Are other students generally engaged? Do you feel that you are treated differently than other students?). Ideally for the mentee, this will draw a picture of whether their displeasure stems from their mentor, research, or a combination of both. The mentee can then take next steps with confidence on what will improve their happiness and engagement within the field. 

The mentor checklist will aim to invoke thought about how the PI interacts with students (i.e., do you feel that you communicate with each of your students equitably? Do you provide requested guidance?) and how they have structured the research group (i.e. are there multiple collaborators your students can turn to? Are students encouraged to work together?). Ideally for the mentor, this will draw a picture of whether they are mentoring well on fundamental principles, if their group dynamic is established in an ethical and effective manner, or if there are deficiencies in the mentorship scheme present. It will also ideally prompt introspection on the inherent power dynamics that exist in any mentor/mentee relationship, regardless of how the mentor views the relationship. 

These checklists will ideally be implemented into the developed mentorship training described above as a measure to constantly check in throughout a mentoring relationship. This will also work in combination with the mentorship compacts that will be encouraged at the beginning of each mentoring relationship.

By focusing on constellation mentorship models and giving students a safe space to grow amongst peers, bolstering physics identity through peer and faculty mentoring, and pointedly confronting and overcoming common barriers in STEM, we can increase the rates of retention and eventually recruitment in physics and astronomy. 

\begin{acknowledgements}

The authors and Spectrum leaders would like to offer their sincerest thanks to the other co-Founders of Spectrum: Dr. Jenna Cann, Carly Solis, and Kathryn Fernandez. The authors and Spectrum leaders would like to thank the Faculty Advisory Board of Spectrum, Dr. Paula Danquah-Brobby, Dr. Fernando Camelli, Dr. Erhai Zhao, Dr. Patrick Vora, and Brooke Vaughn for their constant support of Spectrum and their aid at each step. The authors thank Dr. Ferah Munshi for providing funding. The authors also gratefully acknowledge Dr. Tom Rice and the American Astronomical Society (AAS) for providing the education mini-grant with which this work was made possible.
N. L. gratefully acknowledges financial support from an NSF GRFP. 

\end{acknowledgements}

\section{Appendix}
\label{sec:appendix}
\subsection{Mentoring Compact}

We provide herein the mentoring compact that was developed as a part of the mentorship training. This compact will aid in prompting open dialogue between mentors and mentees, without putting the onus on one party or another to discuss all facets of a successful relationship. The compact will be filled out together, simultaneously, and ideally re-discussed with some consistency. 

\begin{figure*}[]
\centering
\includegraphics[scale=0.71]{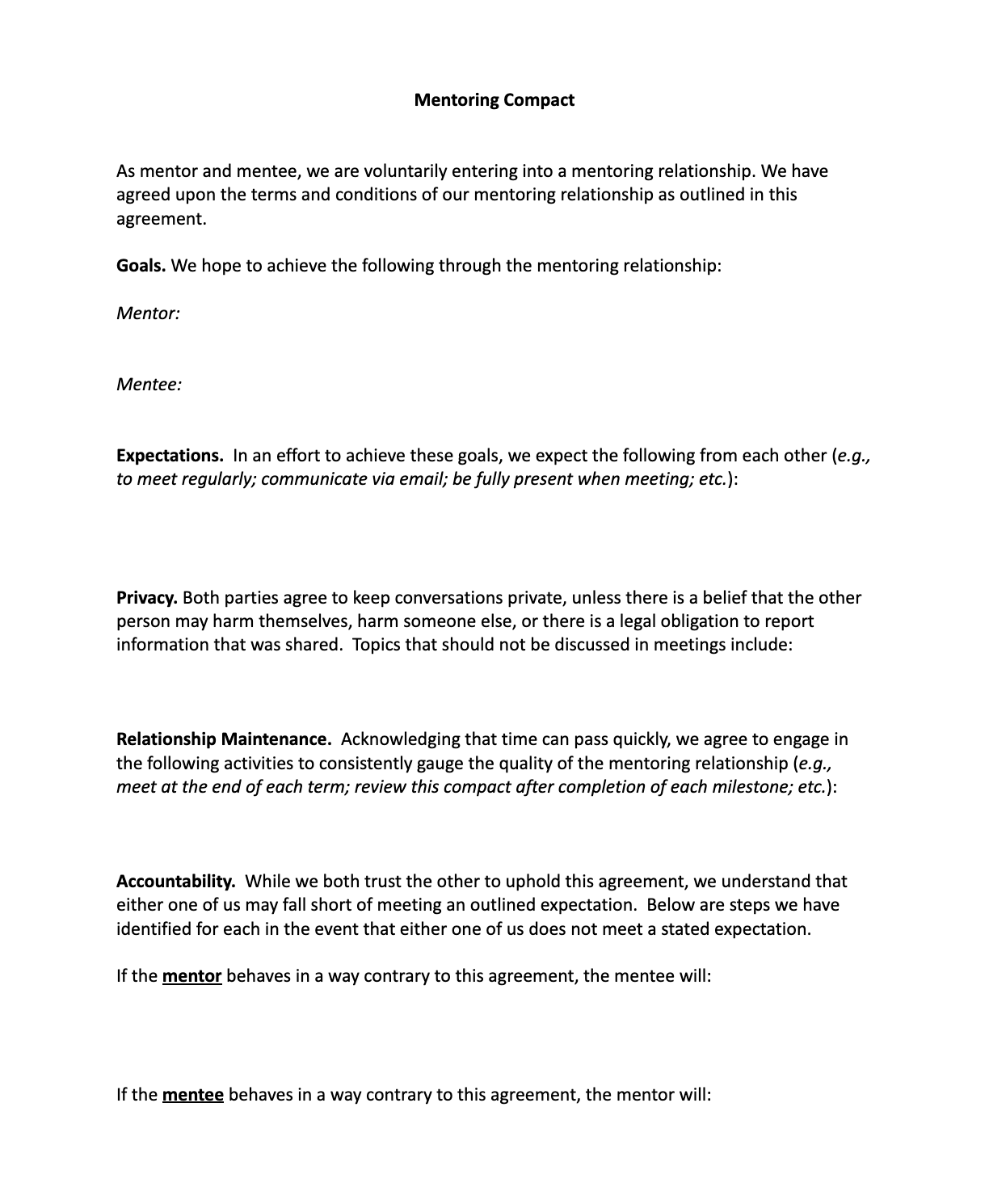}
\caption{Page 1 of the Mentoring Compact Agreement, meant to be filled out by mentors and mentees together.}
\label{fig:mentorcompact1}
\end{figure*} 

\begin{figure*}[]
\centering
\includegraphics[scale=0.7]{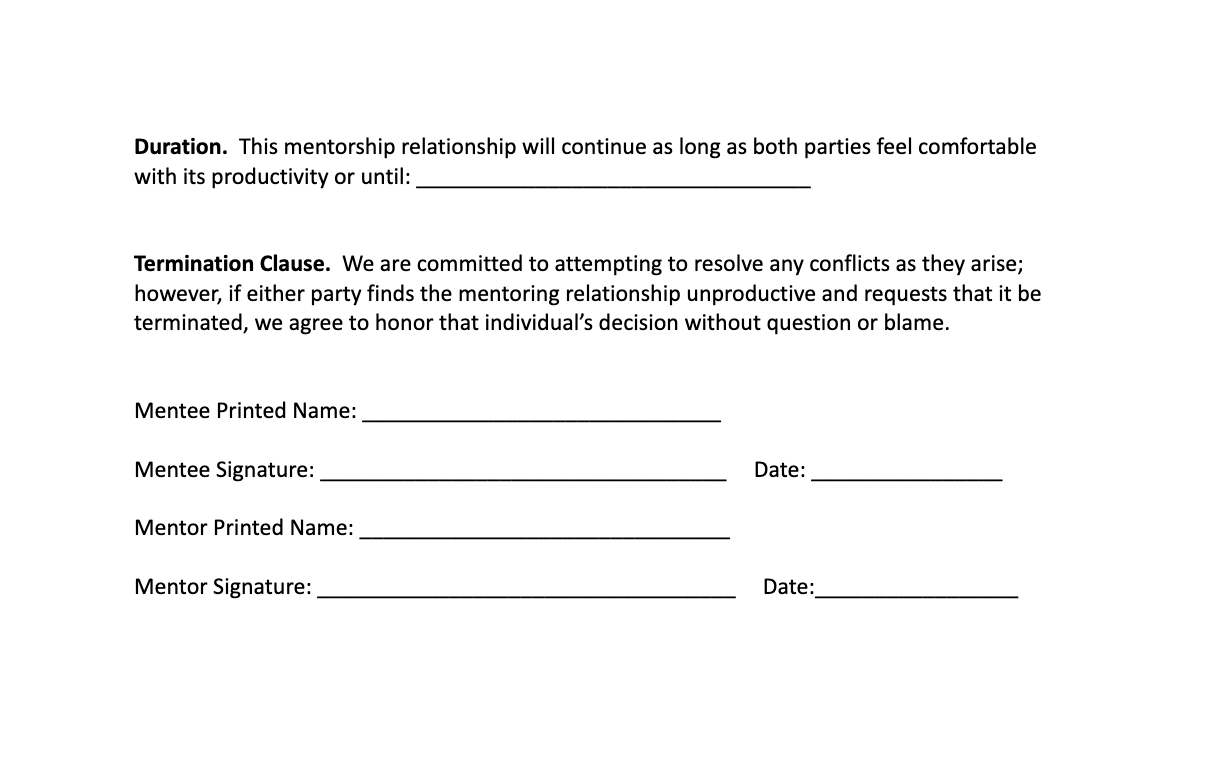}
\caption{Page 2 of the Mentoring Compact Agreement.}
\label{fig:mentorcompact2}
\end{figure*}

\subsection{Faculty Case Studies}

We provide herein the case studies that were developed for faculty specifically to understand their roles in student and mentee lives. These present difficult scenarios that require careful thought before answering, while also covering a number of potential topics (bystander intervention, white knighting, etc.). These case studies are provided, with two minutes to consider, before entering into discussion and gently redirecting to the best approach. 

\begin{figure*}[]
\centering
\includegraphics[scale=0.7]{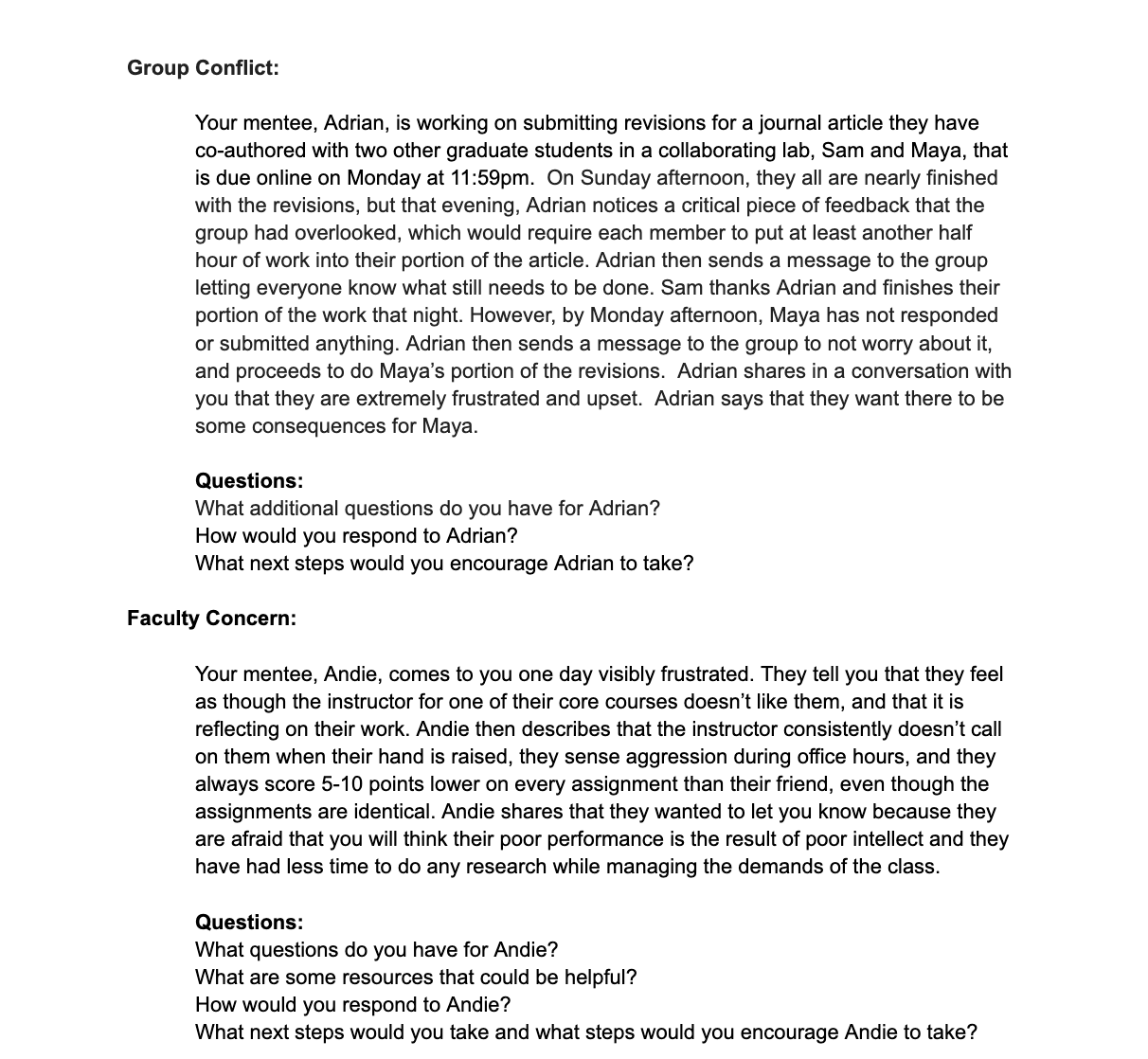}
\caption{The first two provided case studies for faculty, with questions.}
\label{fig:casestudy1}
\end{figure*} 

\begin{figure*}[]
\centering
\includegraphics[scale=0.7]{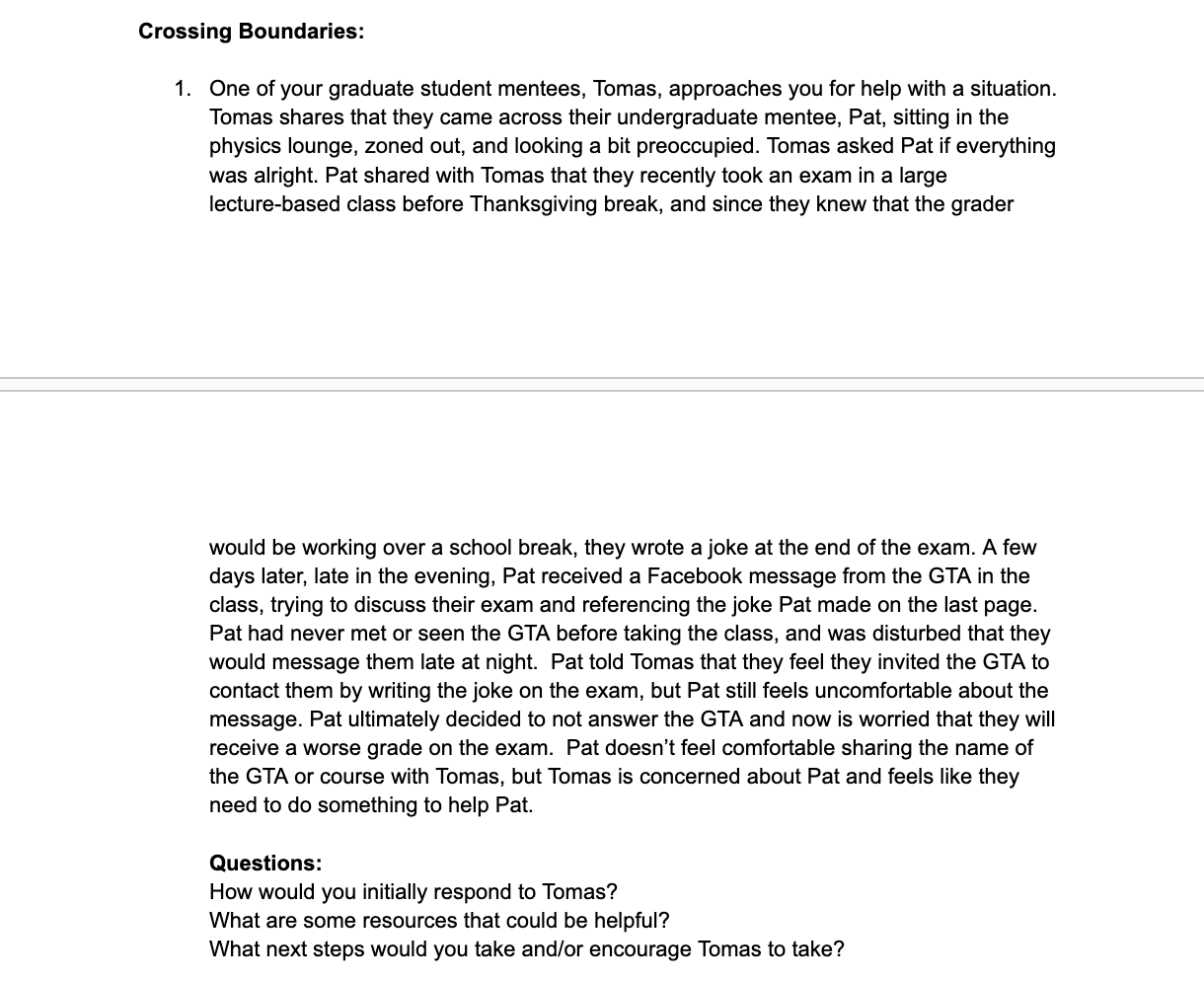}
\caption{The third provided case studies for faculty, with questions.}
\label{fig:casestudy2}
\end{figure*}

\begin{figure*}[]
\centering
\includegraphics[scale=0.7]{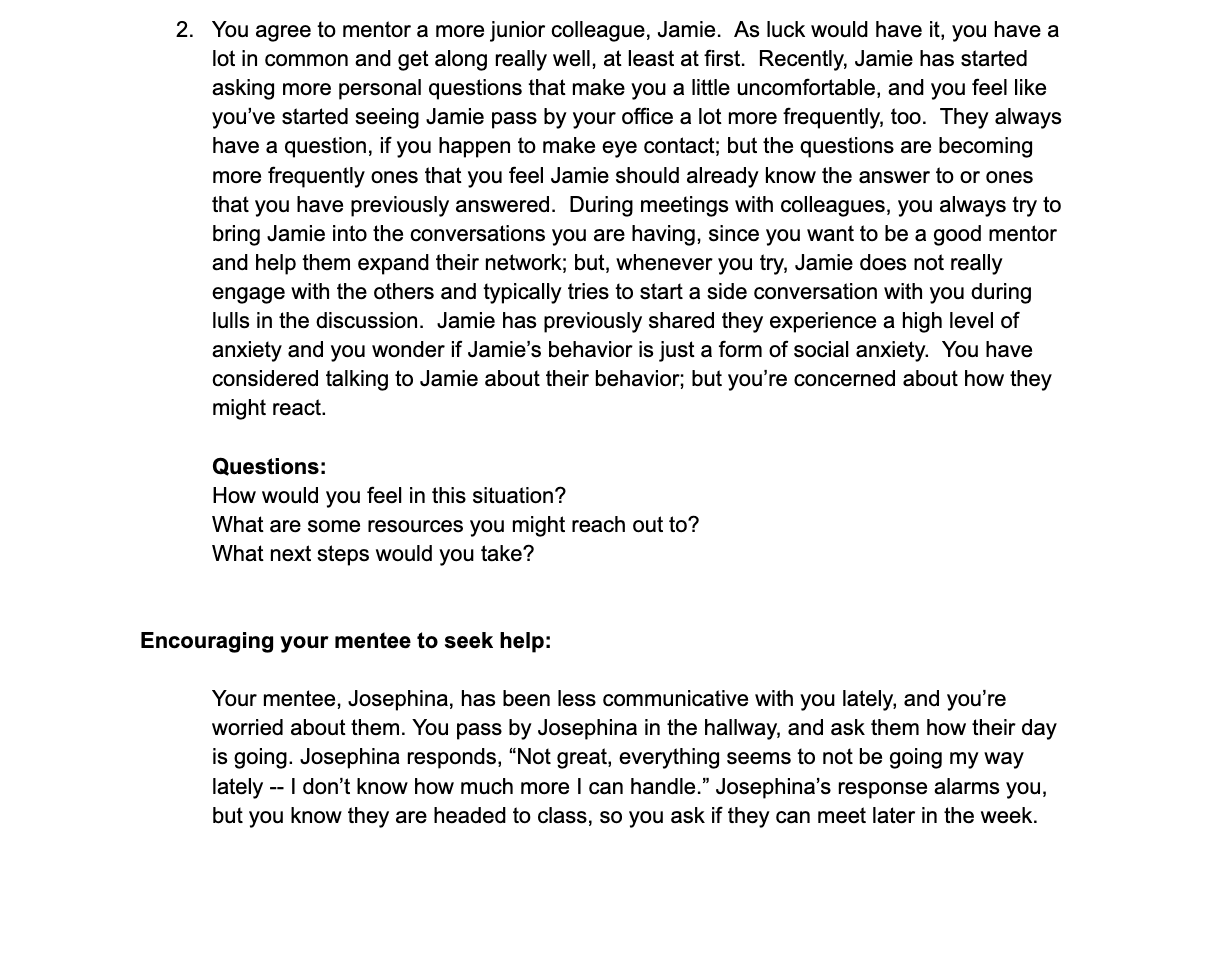}
\caption{The fourth and fifth provided case studies for faculty, with questions.}
\label{fig:casestudy3}
\end{figure*}

\begin{figure*}[]
\centering
\includegraphics[scale=0.7]{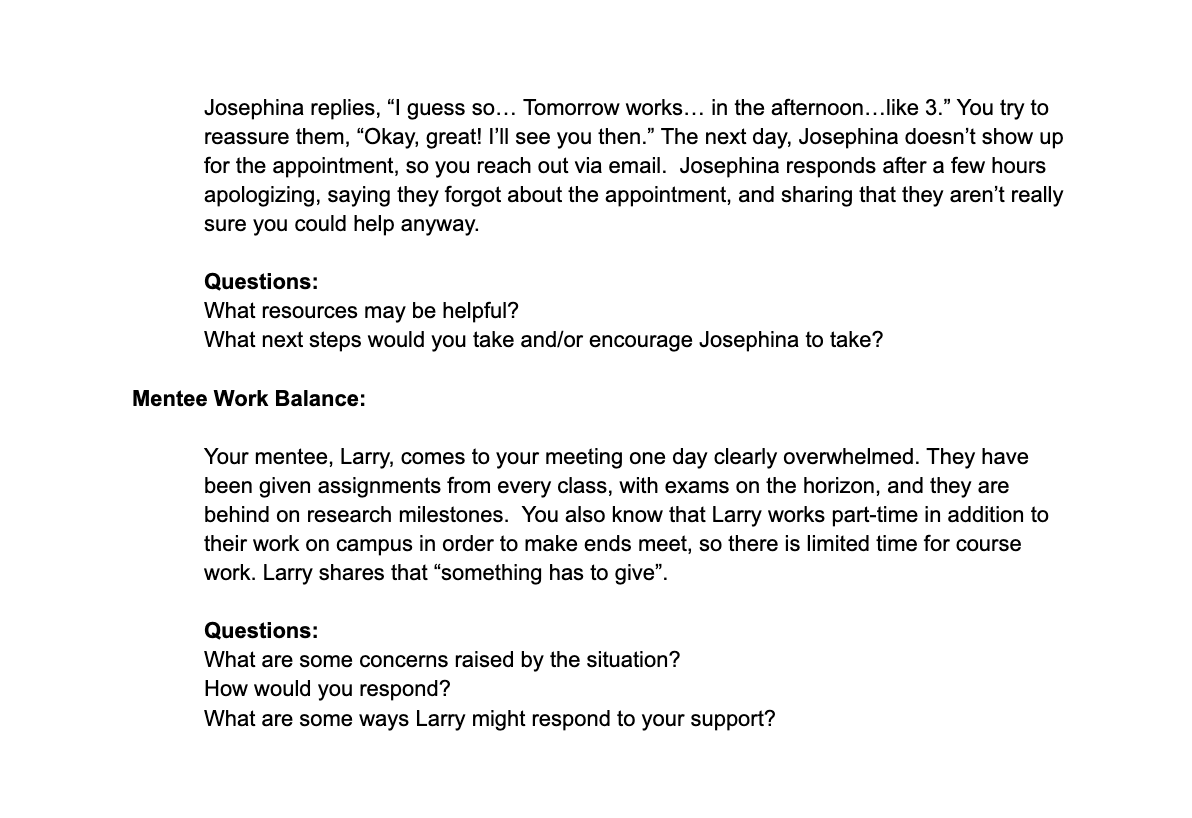}
\caption{The fifth and sixth provided case studies for faculty, with questions.}
\label{fig:casestudy4}
\end{figure*}

\clearpage

\bibliographystyle{aasjournal}
\bibliography{main}{}

\begin{thebibliography}{}
\expandafter\ifx\csname natexlab\endcsname\relax\def\natexlab#1{#1}\fi
\providecommand{\url}[1]{\href{#1}{#1}}
\providecommand{\dodoi}[1]{doi:~\href{http://doi.org/#1}{\nolinkurl{#1}}}
\providecommand{\doeprint}[1]{\href{http://ascl.net/#1}{\nolinkurl{http://ascl.net/#1}}}
\providecommand{\doarXiv}[1]{\href{https://arxiv.org/abs/#1}{\nolinkurl{https://arxiv.org/abs/#1}}}

\bibitem[{Aikens {et~al.}(2017)Aikens, Robertson, Sadselia, Watkins, Evans,
  Runyon, Eby, \& Dolan}]{aikens_race_2017}
Aikens, M.~L., Robertson, M.~M., Sadselia, S., {et~al.} 2017, CBE—Life
  Sciences Education, 16, ar34, \dodoi{10.1187/cbe.16-07-0211}

\bibitem[{Alleyne(2004)}]{alleyne2004black}
Alleyne, A. 2004, Counselling \& Psychotherapy Research, 4, 4,
  \dodoi{10.1080/14733140412331384008}

\bibitem[{Apriceno {et~al.}(2020)Apriceno, Levy, \&
  London}]{apriceno_mentorship_2020}
Apriceno, M., Levy, S.~R., \& London, B. 2020, Journal of College Student
  Development, 61, 643, \dodoi{10.1353/csd.2020.0061}

\bibitem[{Barr-Walker {et~al.}(2020)Barr-Walker, Werner, Kellermeyer, \&
  Bass}]{barr2020coping}
Barr-Walker, J., Werner, D.~A., Kellermeyer, L., \& Bass, M.~B. 2020, Evidence
  Based Library and Information Practice, 15, 24

\bibitem[{Block {et~al.}(2011)Block, Koch, Liberman, Merriweather, \&
  Roberson}]{block_contending_2011}
Block, C.~J., Koch, S.~M., Liberman, B.~E., Merriweather, T.~J., \& Roberson,
  L. 2011, The Counseling Psychologist, 39, 570,
  \dodoi{10.1177/0011000010382459}

\bibitem[{Blondeau \& Awad(2018)}]{blondeau2018relation}
Blondeau, L.~A., \& Awad, G.~H. 2018, Journal of Career Development, 45, 253

\bibitem[{Brown(1990)}]{brown1990racism}
Brown, D. 1990, Virginia Law Review, 295

\bibitem[{Casad \& Bryant(2016)}]{casad2016addressing}
Casad, B.~J., \& Bryant, W.~J. 2016, Frontiers in Psychology, 8

\bibitem[{Chakraverty(2020)}]{chakraverty2020impostor}
Chakraverty, D. 2020, International Journal of Doctoral Studies, 15, 433

\bibitem[{Chan(2008)}]{chan_mentoring_2008}
Chan, A.~W. 2008, Mentoring \& Tutoring: Partnership in Learning, 16, 263,
  \dodoi{10.1080/13611260802231633}

\bibitem[{Clance \& Imes(1978)}]{clance1978imposter}
Clance, P.~R., \& Imes, S.~A. 1978, Psychotherapy: Theory, research \&
  practice, 15, 241

\bibitem[{Cohen \& McConnell(2019)}]{cohen2019fear}
Cohen, E.~D., \& McConnell, W.~R. 2019, The Sociological Quarterly, 60, 457

\bibitem[{Collings {et~al.}(2014)Collings, Swanson, \&
  Watkins}]{collings_impact_2014}
Collings, R., Swanson, V., \& Watkins, R. 2014, Higher Education, 68, 927,
  \dodoi{10.1007/s10734-014-9752-y}

\bibitem[{{Committee on Effective Mentoring in STEMM} {et~al.}(2019){Committee
  on Effective Mentoring in STEMM}, {Board on Higher Education and Workforce},
  {Policy and Global Affairs}, \& {National Academies of Sciences, Engineering,
  and Medicine}}]{byars-winston_science_2019}
{Committee on Effective Mentoring in STEMM}, {Board on Higher Education and
  Workforce}, {Policy and Global Affairs}, \& {National Academies of Sciences,
  Engineering, and Medicine}. 2019, The {Science} of {Effective} {Mentorship}
  in {STEMM}, ed. A.~Byars-Winston \& M.~L. Dahlberg (Washington, D.C.:
  National Academies Press), \dodoi{10.17226/25568}

\bibitem[{{Dancy} \& {Hodari}(2022)}]{dancy22}
{Dancy}, M., \& {Hodari}, A. 2022, arXiv e-prints, arXiv:2210.03522,
  \dodoi{10.48550/arXiv.2210.03522}

\bibitem[{Dennehy \& Dasgupta(2017)}]{dennehy17}
Dennehy, T.~C., \& Dasgupta, N. 2017, Proceedings of the National Academy of
  Sciences, 114, 5964–5969, \dodoi{10.1073/pnas.1613117114}

\bibitem[{Elbogen {et~al.}(2022)Elbogen, Lanier, Griffin, Blakey, Gluff,
  Wagner, \& Tsai}]{elbogen2022national}
Elbogen, E.~B., Lanier, M., Griffin, S.~C., {et~al.} 2022, Cyberpsychology,
  Behavior, and Social Networking, 25, 409

\bibitem[{Evans \& Broido(2002)}]{evans2002experiences}
Evans, N.~J., \& Broido, E.~M. 2002, in Addressing Homophobia and Heterosexism
  on College Campuses, ed. E.~P. Cramer (Binghamton, NY: Harrington Park
  Press), 29--42

\bibitem[{Feenstra {et~al.}(2020)Feenstra, Begeny, Ryan, Rink, Stoker, \&
  Jordan}]{feenstra2020contextualizing}
Feenstra, S., Begeny, C.~T., Ryan, M.~K., {et~al.} 2020, Frontiers in
  psychology, 11, 3206

\bibitem[{Ferrari \& Thompson(2006)}]{ferrari2006impostor}
Ferrari, J.~R., \& Thompson, T. 2006, Personality and Individual Differences,
  40, 341

\bibitem[{Fisher {et~al.}(2019)Fisher, Mendoza-Denton, Patt, Young, Eppig,
  Garrell, Rees, Nelson, \& Richards}]{fisher2019structure}
Fisher, A.~J., Mendoza-Denton, R., Patt, C., {et~al.} 2019, PloS one, 14,
  e0209279

\bibitem[{Flaxman {et~al.}(1988)}]{flaxman1988youth}
Flaxman, E., {et~al.} 1988

\bibitem[{Fraenza(2016)}]{fraenza2016role}
Fraenza, C.~B. 2016, Online Learning, 20, 230

\bibitem[{Godwin {et~al.}(2016)Godwin, Potvin, Hazari, \& Lock}]{godwin16}
Godwin, A., Potvin, G., Hazari, Z., \& Lock, R. 2016, Journal of Engineering
  Education, 105, 312–340, \dodoi{10.1002/jee.20118}

\bibitem[{Haeger \& Fresquez(2016)}]{haeger2016mentoring}
Haeger, H., \& Fresquez, C. 2016, CBE—Life Sciences Education, 15, ar36

\bibitem[{Harris \& Lee(2019)}]{harris_advocate-mentoring_2019}
Harris, T.~M., \& Lee, C.~N. 2019, Communication Education, 68, 103,
  \dodoi{10.1080/03634523.2018.1536272}

\bibitem[{Harvey \& Katz(1985)}]{harvey1985if}
Harvey, J.~C., \& Katz, C. 1985, If I'm so successful, why do I feel like a
  fake?: The impostor phenomenon (St. Martin's Press New York)

\bibitem[{{Hazari} {et~al.}(2010){Hazari}, {Sonnert}, {Sadler}, \&
  {Shanahan}}]{hazari10}
{Hazari}, Z., {Sonnert}, G., {Sadler}, P.~M., \& {Shanahan}, M.-C. 2010,
  Journal of Research in Science Teaching, \dodoi{10.1002/tea.20363}

\bibitem[{Inzlicht \& Good(2006)}]{inzlicht2006environments}
Inzlicht, M., \& Good, C. 2006, in Stigma and group inequality (Psychology
  Press), 143--164

\bibitem[{Ito \& McPherson(2018)}]{ito2018factors}
Ito, T.~A., \& McPherson, E. 2018, Frontiers in psychology, 9, 400169

\bibitem[{Johnson \& Elliott(2020)}]{johnson2020culturally}
Johnson, A., \& Elliott, S. 2020, Journal of microbiology \& biology education,
  21, 10

\bibitem[{Jones(2018)}]{jones18}
Jones, N. 2018, 562, \dodoi{10.1038/D41586-018-06834-Y}

\bibitem[{Joshi {et~al.}(2019)Joshi, Aikens, \& Dolan}]{joshi_direct_2019}
Joshi, M., Aikens, M.~L., \& Dolan, E.~L. 2019, BioScience, 69, 389,
  \dodoi{10.1093/biosci/biz039}

\bibitem[{Kendall(2003)}]{kendall2003}
Kendall, F.~E. 2003, How to Be an Ally if You Are a Person with Privilege,
  \url{http://www.scn.org/friends/ally.html}

\bibitem[{Kim \& Sinatra(2018)}]{kim2018science}
Kim, A.~Y., \& Sinatra, G.~M. 2018, International journal of STEM education, 5,
  1

\bibitem[{L.~Baker {et~al.}(2014)L.~Baker, J.~Pifer, \&
  A.~Griffin}]{l_baker_mentor-protege_2014}
L.~Baker, V., J.~Pifer, M., \& A.~Griffin, K. 2014, International Journal for
  Researcher Development, 5, 83, \dodoi{10.1108/IJRD-04-2014-0003}

\bibitem[{Locks {et~al.}(2008)Locks, Hurtado, Bowman, \&
  Oseguera}]{locks_extending_2008}
Locks, A.~M., Hurtado, S., Bowman, N.~A., \& Oseguera, L. 2008, The Review of
  Higher Education, 31, 257, \dodoi{10.1353/rhe.2008.0011}

\bibitem[{Markle {et~al.}(2022)Markle, Williams, Williams, deGravelles,
  Bagayoko, \& Warner}]{markle_supporting_nodate}
Markle, R.~S., Williams, T.~M., Williams, K.~S., {et~al.} 2022, 7, 674669

\bibitem[{Maton {et~al.}(2000)Maton, Hrabowski~III, \&
  Schmitt}]{maton2000african}
Maton, K.~I., Hrabowski~III, F.~A., \& Schmitt, C.~L. 2000, Journal of Research
  in Science Teaching: The Official Journal of the National Association for
  Research in Science Teaching, 37, 629

\bibitem[{Meador(2018)}]{meador2018examining}
Meador, A. 2018, School Science and Mathematics, 118, 61

\bibitem[{Miller(2019)}]{miller19}
Miller, J.~V. 2019, Physics Today, \dodoi{10.1063/PT.3.31335}

\bibitem[{Mondisa \& McComb(2015)}]{mondisa_social_2015}
Mondisa, J.-L., \& McComb, S.~A. 2015, Mentoring \& Tutoring: Partnership in
  Learning, 23, 149, \dodoi{10.1080/13611267.2015.1049018}

\bibitem[{{National Academies of Sciences} \& Medicine(2021)}]{decadal20}
{National Academies of Sciences}, E., \& Medicine. 2021, {Pathways to Discovery
  in Astronomy and Astrophysics for the 2020s}, \dodoi{10.17226/26141}

\bibitem[{{Pando}(2022)}]{pando22}
{Pando}, J. 2022, in APS Meeting Abstracts, Vol. 2022, APS April Meeting
  Abstracts, B06.001

\bibitem[{Pfund {et~al.}(2016)Pfund, Byars-Winston, Branchaw, Hurtado, \&
  Eagan}]{pfund_defining_2016}
Pfund, C., Byars-Winston, A., Branchaw, J., Hurtado, S., \& Eagan, K. 2016,
  AIDS and Behavior, 20, 238, \dodoi{10.1007/s10461-016-1384-z}

\bibitem[{Pfund {et~al.}(2006)Pfund, Maidl~Pribbenow, Branchaw, Miller~Lauffer,
  \& Handelsman}]{pfund_merits_2006}
Pfund, C., Maidl~Pribbenow, C., Branchaw, J., Miller~Lauffer, S., \&
  Handelsman, J. 2006, Science, 311, 473, \dodoi{10.1126/science.1123806}

\bibitem[{{Porter} \& {Ivie}(2019)}]{ivie19}
{Porter}, A.~M., \& {Ivie}, R. 2019, Women in Physics and Astronomy: 2019.
\newblock
  \url{https://www.aip.org/statistics/reports/women-physics-and-astronomy-2019}

\bibitem[{Pruitt \& Isaac(1985)}]{pruitt_discrimination_1985}
Pruitt, A.~S., \& Isaac, P.~D. 1985, The Journal of Negro Education, 54, 526,
  \dodoi{10.2307/2294713}

\bibitem[{Robnett {et~al.}(2019)Robnett, Nelson, Zurbriggen, Crosby, \&
  Chemers}]{robnett2019form}
Robnett, R.~D., Nelson, P.~A., Zurbriggen, E.~L., Crosby, F.~J., \& Chemers,
  M.~M. 2019, Emerging Adulthood, 7, 180

\bibitem[{San~Miguel \& Kim(2015)}]{san_miguel_successful_2015}
San~Miguel, A.~M., \& Kim, M.~M. 2015, Journal of Career Development, 42, 133,
  \dodoi{10.1177/0894845314542248}

\bibitem[{Sanford {et~al.}(2015)Sanford, Ross, Blake, \&
  Cambiano}]{sanford2015finding}
Sanford, A.~A., Ross, E. M. R.~M., Blake, S.~J., \& Cambiano, R.~L. 2015,
  Advancing Women in Leadership Journal, 35, 31

\bibitem[{Stachl \& Baranger(2020)}]{stachl2020sense}
Stachl, C.~N., \& Baranger, A.~M. 2020, PloS one, 15, e0233431

\bibitem[{Steele \& Aronson(1995)}]{steele1995stereotype}
Steele, C.~M., \& Aronson, J. 1995, Journal of personality and social
  psychology, 69, 797

\bibitem[{Stolle-McAllister {et~al.}(2011)Stolle-McAllister, Sto.~Domingo, \&
  Carrillo}]{stolle-mcallister_meyerhoff_2011}
Stolle-McAllister, K., Sto.~Domingo, M.~R., \& Carrillo, A. 2011, Journal of
  Science Education and Technology, 20, 5, \dodoi{10.1007/s10956-010-9228-5}

\bibitem[{Sue(2010)}]{sue2010microaggressions}
Sue, D.~W. 2010, Microaggressions in Everyday Life: Race, Gender, and Sexual
  Orientation (Hoboken, NJ: Wiley)

\bibitem[{Swim {et~al.}(2001)Swim, Hyers, Cohen, \&
  Ferguson}]{swim2001everyday}
Swim, J.~K., Hyers, L.~L., Cohen, L.~L., \& Ferguson, M.~J. 2001, Journal of
  Social Issues, 57, 31, \dodoi{10.1111/0022-4537.00200}

\bibitem[{Tatum(2000)}]{tatum2000whoami}
Tatum, B. 2000, in Readings for Diversity and Social Justice: An Anthology on
  Racism, Antisemitism, Sexism, Heterosexism, Ableism, and Classism, ed.
  M.~Adams, W.~J. Blumenfeld, R.~Castaneda, H.~Hackman, M.~Peters, \&
  X.~Zuñiga (New York, NY: Routledge), 39--44

\bibitem[{Thelamour {et~al.}(2019)Thelamour, George~Mwangi, \&
  Ezeofor}]{thelamour2019we}
Thelamour, B., George~Mwangi, C., \& Ezeofor, I. 2019, Journal of Diversity in
  Higher Education, 12, 266

\bibitem[{Thomas \& Erdei(2018)}]{thomas2018stemming}
Thomas, N., \& Erdei, R. 2018, in 2018 CoNECD-The Collaborative Network for
  Engineering and Computing Diversity Conference

\bibitem[{Thompson {et~al.}(1998)Thompson, Davis, \&
  Davidson}]{thompson1998attributional}
Thompson, T., Davis, H., \& Davidson, J. 1998, Personality and Individual
  differences, 25, 381

\bibitem[{Thompson {et~al.}(2000)Thompson, Foreman, \&
  Martin}]{thompson2000impostor}
Thompson, T., Foreman, P., \& Martin, F. 2000, Personality and Individual
  differences, 29, 629

\bibitem[{{Trawick}~{Monroe-White} {et~al.}(2021){Trawick}~{Monroe-White},
  {Joseph}, {Tolbert}, {Tola}, \& {Haynes}}]{trawick21}
{Trawick}~{Monroe-White}, T., {Joseph}, M., {Tolbert}, N., {Tola}, J., \&
  {Haynes}, J. 2021, \dodoi{doi:10.1177/23733799211049238}

\bibitem[{Woodcock {et~al.}(2012)Woodcock, Hernandez, Estrada, \&
  Schultz}]{woodcock2012consequences}
Woodcock, A., Hernandez, P.~R., Estrada, M., \& Schultz, P. 2012, Journal of
  personality and social psychology, 103, 635

\bibitem[{{Woods} \& {Walker}(2022)}]{walker22}
{Woods}, P., \& {Walker}, A. 2022, Nature Astronomy, 6, 622,
  \dodoi{10.1038/s41550-022-01704-0}

\bibitem[{Young {et~al.}(2015)Young, Anderson, \&
  Stewart}]{young_hierarchical_2015}
Young, K., Anderson, M., \& Stewart, S. 2015, Journal of Diversity in Higher
  Education, 8, 61, \dodoi{10.1037/a0038464}

\bibitem[{Zaniewski \& Reinholz(2016)}]{zaniewski2016increasing}
Zaniewski, A.~M., \& Reinholz, D. 2016, International journal of STEM
  education, 3, 1

\end{thebibliography}

\end{document}